\documentclass[sigconf, nonacm]{acmart}

\usepackage{balance}
\usepackage{xcolor}
\usepackage{placeins}
\newcommand\ourmethod{TOPJoin}
\newcommand\nextiaS{Nextia$_{JD}$-S }
\newcommand\nextiaM{Nextia$_{JD}$-M }
\newcommand\nextia{Nextia$_{JD}$ }
\newcommand{\best}[1]{\textbf{#1}}
\newcommand{\secondbest}[1]{\underline{#1}}

\usepackage{subcaption}
\usepackage[capitalise,nameinlink]{cleveref}
\usepackage{graphicx}
\usepackage{algorithm}
\usepackage{algpseudocode}
\usepackage{multirow}
\usepackage{xcolor}
\usepackage{pgfplots}
\usepackage{tikz}

\usepackage{pgfplots}
\usepackage{tikz}
\usepgfplotslibrary{groupplots,dateplot}
\usetikzlibrary{patterns, shapes.arrows, patterns.meta, plotmarks}
\usetikzlibrary{external}
\usetikzlibrary{er,positioning}
\newcommand{\inputtikz}[1]{%
\resizebox{\columnwidth}{!}{
  \tikzsetnextfilename{#1}%
  \includegraphics{#1.pdf}%
}
}

\usepackage[framemethod=default]{mdframed}

\usepackage{colortbl}

\definecolor{bgcolor}{RGB}{240,240,240}
\newmdenv[hidealllines=true,backgroundcolor=bgcolor,innerleftmargin=8pt,innerrightmargin=8pt,innertopmargin=0.1em,innerbottommargin=0.5em,skipbelow=\baselineskip,skipabove=\baselineskip]{callout}

\usepackage{ifthen}

\newcommand{\introparagraph}[1]{\paragraph{#1}} %

\acmConference{}{}{}

\begin{document}
\newboolean{istr}
\setboolean{istr}{true}

\title{
[Experiments \& Analysis]  \\
Evaluating Joinable Column Discovery Approaches for Context-Aware Search
}

\author{Harsha Kokel}
\affiliation{
\institution{IBM Research}
\state{New York}
\country{USA}
}

\author{Aamod Khatiwada}
\authornote{Work done while at IBM Research.}
\affiliation{
\institution{Northeastern University}
\city{Boston}
\state{Massachusetts}
\country{USA}
}

\author{Tejaswini Pedapati}
\affiliation{
\institution{IBM Research}
\state{New York}
\country{USA}
}

\author{Haritha Ananthakrishnan}
\affiliation{
\institution{IBM Research}
\state{New York}
\country{USA}
}

\author{Oktie Hassanzadeh}
\affiliation{
\institution{IBM Research}
\state{New York}
\country{USA}
}

\author{Horst Samulowitz}
\affiliation{
\institution{IBM Research}
\state{New York}
\country{USA}
}

\author{Kavitha Srinivas}
\affiliation{
\institution{IBM Research}
\state{New York}
\country{USA}
}

\renewcommand{\shortauthors}{Kokel et al.}

\begin{abstract}
Joinable Column Discovery is a critical challenge in automating enterprise data analysis. While existing approaches focus on syntactic overlap and semantic similarity, there remains limited understanding of which methods perform best for different data characteristics and how multiple criteria influence discovery effectiveness. We present a comprehensive experimental evaluation of joinable column discovery methods across diverse scenarios. Our study compares syntactic and semantic techniques on seven benchmarks covering relational databases and data lakes. We analyze six key criteria—unique values, intersection size, join size, reverse join size, value semantics, and metadata semantics—and examine how combining them through ensemble ranking affects performance. Our analysis reveals differences in method behavior across data contexts and highlights the benefits of integrating multiple criteria for robust join discovery. We provide empirical evidence on when each criterion matters, compare pre-trained embedding models for semantic joins, and offer practical guidelines for selecting suitable methods based on dataset characteristics. Our findings show that metadata and value semantics are crucial for data lakes, size-based criteria play a stronger role in relational databases, and ensemble approaches consistently outperform single-criterion methods.
\end{abstract}

\maketitle

\section{Introduction}

With the rapid growth of enterprise data repositories, automating data discovery has become a central challenge in enabling scalable analytics. Data discovery refers to the process of identifying relevant data sources that can answer specific analytical or business questions. In the context of enterprise applications and natural language interfaces for data analysis~\cite{WeideleRBVBCMMA23,Mihindukulasooriya23,WeideleMVRSFAMA24}, a key component of this process is the discovery of \emph{joinable tables}: given a query table and a designated join column, determine which other tables and columns can be joined to enrich the query table with additional attributes and insights~\cite{ALITE_khatiwadaSGM22}.

Traditional approaches to joinable table discovery primarily rely on measuring similarity between columns. Two columns are considered \emph{syntactically joinable} when they share overlapping values~\cite{JOSIE_ZhuDNM19,LSHENSEMBLE_ZhuNPM16}, and \emph{semantically joinable} when their embeddings are close in a semantic vector space~\cite{PEXESO_DongT0O21,DeepJoin_Dong0NEO23,WarpGate_CongGFJD23}. While these approaches have shown promising results, they also exhibit significant limitations when applied to heterogeneous enterprise data lakes that contain tables from diverse and often unrelated domains. In such settings, naive similarity-based joins may lead to numerous false positives—joins that are statistically similar but contextually irrelevant—thereby reducing both precision and interpretability of downstream analyses.

Despite several recent advances in data discovery and table search~\cite{DBLP:conf/icde/SantosBMF22,DBLP:journals/pvldb/DengCCYCYSWLCJZJZWYWT24,DBLP:journals/corr/abs-2310-02656,DBLP:journals/tkde/GuoMHCG25,ChristensenLeventidisLissandriniDiRoccoMillerHose2025}, there remains limited understanding of \emph{when} particular methods perform well, \emph{why} they fail, and how different similarity criteria influence the effectiveness of join discovery. 
While these recent systems have introduced new models, indices, and benchmarks for large-scale or semantic data discovery, they typically focus on algorithmic innovation rather than comparative insight. 
In this paper, we take an empirical and analytical perspective, conducting a systematic experimental study of the criteria used for joinable column discovery to provide a clear, data-driven assessment of the accuracy of various methods across diverse settings.

To this end, we consider both syntactic methods (e.g., LSH Ensemble~\cite{LSHENSEMBLE_ZhuNPM16}, JOSIE~\cite{JOSIE_ZhuDNM19}) and semantic methods (e.g., DeepJoin~\cite{DeepJoin_Dong0NEO23}, WarpGate~\cite{WarpGate_CongGFJD23}), and analyze how their performance varies under different data characteristics. We identify six key criteria that govern join effectiveness—\emph{unique values, intersection size, join size, reverse join size, value semantics,} and \emph{metadata semantics}—and study how each contributes to identifying meaningful joins. We further explore how combining these criteria through ensemble ranking can balance trade-offs between precision and recall, yielding more robust behavior across domains.

Our evaluation spans seven benchmarks covering relational databases, enterprise data lakes, and fuzzy-join scenarios. The collection includes both established datasets from prior work and newly curated benchmarks, including manually annotated open data tables designed to capture context-aware joinability. This diversity allows us to assess method behavior across controlled and realistic conditions.

Our study leads to several practical insights. We observe that no single approach dominates across all settings; metadata and value semantics emerge as the most decisive factors for heterogeneous data lakes, while size-based criteria play a stronger role in structured relational databases. Ensemble approaches that integrate multiple criteria consistently outperform individual measures, demonstrating the importance of multi-faceted evaluation in joinable column discovery.

In summary, the contributions of this paper are as follows:
\begin{itemize}
    \item We conduct the first comprehensive experimental analysis of the accuracy of joinable column discovery methods across relational, data lake, and fuzzy-join scenarios.
    \item We introduce a systematic framework for evaluating six key criteria that influence joinability and analyze their individual and combined impact on discovery effectiveness.
    \item We curate and release seven benchmarks, including both prior and newly annotated datasets, to facilitate reproducible and context-aware evaluation.
    \item We provide empirical guidelines for selecting appropriate methods and criteria based on dataset characteristics and analytical objectives.
\end{itemize}

\section{Task Definition and Joinability Criteria}
\label{sec:problem_definition}

Discovering joinable tables is a fundamental step in data discovery and integration, enabling users to enrich a query table with related attributes from other tables. Understanding what makes two columns joinable, and under what conditions certain definitions of joinability are more effective, is central to our analysis. In this section, we formalize the task of joinable column discovery and describe the different factors—or criteria—that influence whether two columns should be considered joinable in practice. These definitions and criteria form the conceptual foundation for our experimental study.

\subsection{Joinable Column Discovery Task}

Two tables can be joined if each of them contains a column that can be joined.
Consequently, the task of discovering joinable tables can be formulated as a \emph{joinable column search problem} defined as follows:

\begin{definition}\label{def:join_search}
    \textbf{Joinable Column Search Problem:} Given a collection of columns $\mathcal{C}$, a query column $q_Q$ from a table $Q$, and a constant $k$, find top-$k$ columns from the collection that are \emph{``joinable''} to the query column $q_Q$.
\end{definition}

Various notions of \emph{``joinable''} columns have been explored 
in the literature~\cite{datalake_tutorial_NargesianZMPA19}. 
Popular definitions are based on set similarity, which measures set intersection size~\cite{JOSIE_ZhuDNM19}, Jaccard similarity~\cite{Mannheim_LehmbergRRMPB15}, or set containment~\cite{LSHENSEMBLE_ZhuNPM16} between a pair of columns. 
Approaches that use such syntactic join measures are often referred to as \emph{equality join} or \emph{equi-join} approaches. 
Recent work has also explored the notion of \emph{semantically joinable}. 
Specifically, Dong et al.~\cite{PEXESO_DongT0O21} defined semantic joinability as a measure of similarity between the embeddings of values in the column (referred to as vector matching). 
Cong et al.~\cite{WarpGate_CongGFJD23} studied semantic column joinability defined as similarity between column embeddings. 
Koutras et al.~\cite{Valentine_KoutrasSIPBFLBK21} defined column pairs as semantically joinable if the columns share semantically equivalent instances. 
Notably, none of these approaches consider table context in the joinable column definition. 
In our study, we focus on joinable columns that are meaningful within their broader table context and refer to these as \emph{context-aware joinable columns}.

\begin{definition}
    \textbf{Context-aware Joinable Columns:} Given a query column $q_Q$ and a target column $t_T$, the following two conditions must be met for the columns to be context-aware joinable:
        (i) The tables $Q$ and $T$ must possess a semantic relationship or belong to related domains, indicating that the data in these tables are connected or relevant to one another. 
        (ii) The columns $q_Q$ and $t_T$ must be semantically joinable, meaning that they can be combined to produce meaningful results that satisfy the query's objectives.
\end{definition}

Consider the following example from Open Data, which is widely used for data discovery~\cite{datalake_tutorial_NargesianZMPA19}. 

\begin{callout}
\begin{example}\label{ex:context_aware_join}
\noindent
\cref{fig:topjoin_main}(a,b)
are about the child population in Texas,~\cref{fig:topjoin_main}(c) is about tobacco retailers in Texas, while ~\cref{fig:topjoin_main}(d) is about Missouri. 
Assume Table I(a) and Column \texttt{County} (denoted as \texttt{a.County}) is the query for joinable column search. 
Typical joinability criteria based on value and semantic overlaps identify 
Tables \texttt{b.County}, \texttt{c.County}, and \texttt{d.County} to be joinable with the query column.
However, from the user's perspective, joining Tables (a) and (b) makes the most sense as it augments the query table to provide insights
such as what proportion of the child population received assistance. 
Joining Tables (a) and (c) may be less preferable because the join will increase the size of the original query table (each row would get repeated for each retailer in that county). 
Joining Tables (a) and (d) does not yield additional insights; in fact, the join is not sensible because the places are not the same. 
So, labeling Table (d) as joinable does not aid in data discovery. 
Furthermore, the representation of the joinable values could be different
(e.g., \verb+Madison+ in Table (a) vs.\ \verb+madison+ in Table (b), or \verb+NY+ vs.\ \verb+New York+). 
Ideally, any computation of joinability should consider such fuzzy joins as well.
\end{example}
\end{callout}

\begin{table*}[t]
\centering
 \caption{Snippets of tables from Open Data.}
    \includegraphics[width=\textwidth]{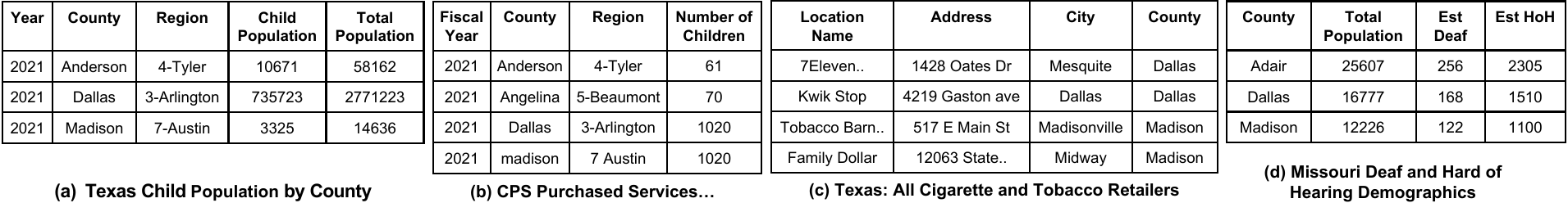}
\label{fig:topjoin_main}
\end{table*}

\subsection{Joinability Criteria}

In the following, we identify key factors that determine whether two columns should be considered contextually joinable and describe how these factors can be estimated to improve search effectiveness and utility in extracting insights.

\paragraph{Unique Values.}
Generally, in databases, we join a
primary key column with a foreign key column. 
A column with a higher percentage of unique values has a
higher probability of being a primary key column. 
In the context of data lakes—where explicit database schemas are often unavailable—the ratio of distinct values in a column to
the total number of rows serves as a useful proxy for assessing uniqueness. 

\paragraph{Intersection Size}
The most basic requirement for joining
two columns is that there must be overlapping values to join on. 
Thus, intersection size is one of the most important criteria for join discovery. 
Intersection size can be computed using exact matches from posting lists. 
However, as it is computationally prohibitive to compute the overlap
between columns in large enterprise data lakes, each column can instead be represented by its MinHash values, and the intersection size can be determined by the Hamming distance between these representations~\cite{LSHForest_BawaCG05}. 
The intersection size can also be normalized to account for the size of the columns. 
Jaccard and containment scores are heavily used in the literature to capture intersection size.

\paragraph{Join Sizes and Reverse Join Sizes}
If we combine~\cref{fig:topjoin_main}(a) and (b), the overall cardinality of the joined table is similar to the original tables and provides additional insights. 
However, if we combine~\cref{fig:topjoin_main}(b) and (c), the cardinality of the resulting table becomes much larger than the original table sizes. 
Furthermore, the columns \texttt{Region}, \texttt{Child Population}, and \texttt{Total Population} in the resulting join table would be duplicated across rows. 
In an ideal join scenario, we want the cardinality of the joined table to be reasonable and to avoid rows where some column values are duplicated. 
Given a query and a candidate table, \textbf{join size} is the size of the combined table when a left join is performed, and \textbf{reverse join size} is the size of the combined table when a right join is performed. 
While exact join size computation is expensive, various approaches have been proposed to estimate the cardinality of the joined table in the literature~\cite{JoinSize_Swami94}. 

\paragraph{Value Semantics}
To understand whether the query column and the candidate column are joinable, the most common strategy is to calculate the intersection size mentioned above. 
If there is a high overlap among them, the columns are deemed joinable. 
However, considering only raw values or their MinHash representations fails to capture the conceptual similarity between columns. 
For instance, two columns with values \textbf{New York, Los Angeles, San Francisco, Washington D.C.} and \textbf{NYC, LA, SF, DC} would have no overlap but are conceptually identical. 
To incorporate the semantics of each column, one can compute the cosine similarity between embeddings of their most frequent values. 
This approach is particularly informative for string columns, where value semantics carry contextual meaning, but is generally less effective for numeric columns.

\paragraph{Disjoint Value Semantics}
Value semantics considers the concepts of the entire column. 
If the query column and the candidate column share many values, their concepts will naturally be similar. 
For example, consider a column containing fruit names: \textbf{Apple, Orange, Blackberry, Pear}, and another containing company names: \textbf{Apple, Oracle, Blackberry, Dell}. 
The shared values \textbf{Apple} and \textbf{Blackberry} lead to high semantic similarity even though in one column they refer to fruits and in the other to technology companies. 
The non-overlapping values (\textbf{Orange} and \textbf{Oracle}) show low semantic similarity, helping to distinguish distinct conceptual domains despite superficial overlap. 
High semantic similarity among disjoint values indicates that columns are indeed related. 
Like value semantics, this criterion is also more effective for string columns.

\paragraph{Metadata Semantics}
Finally, beyond column values, table metadata can be a significant indicator of column relevance. 
This includes table descriptions, dataset tags, column names, and column descriptions. 
As highlighted in Example~\ref{ex:context_aware_join}, the context of the query is an important factor in ensuring meaningful results.

\section{Datasets}
\label{sec:datasets}
\ifthenelse{\boolean{istr}}
{

\begin{figure*}[t]
    \centering
    \includegraphics[scale=0.3]{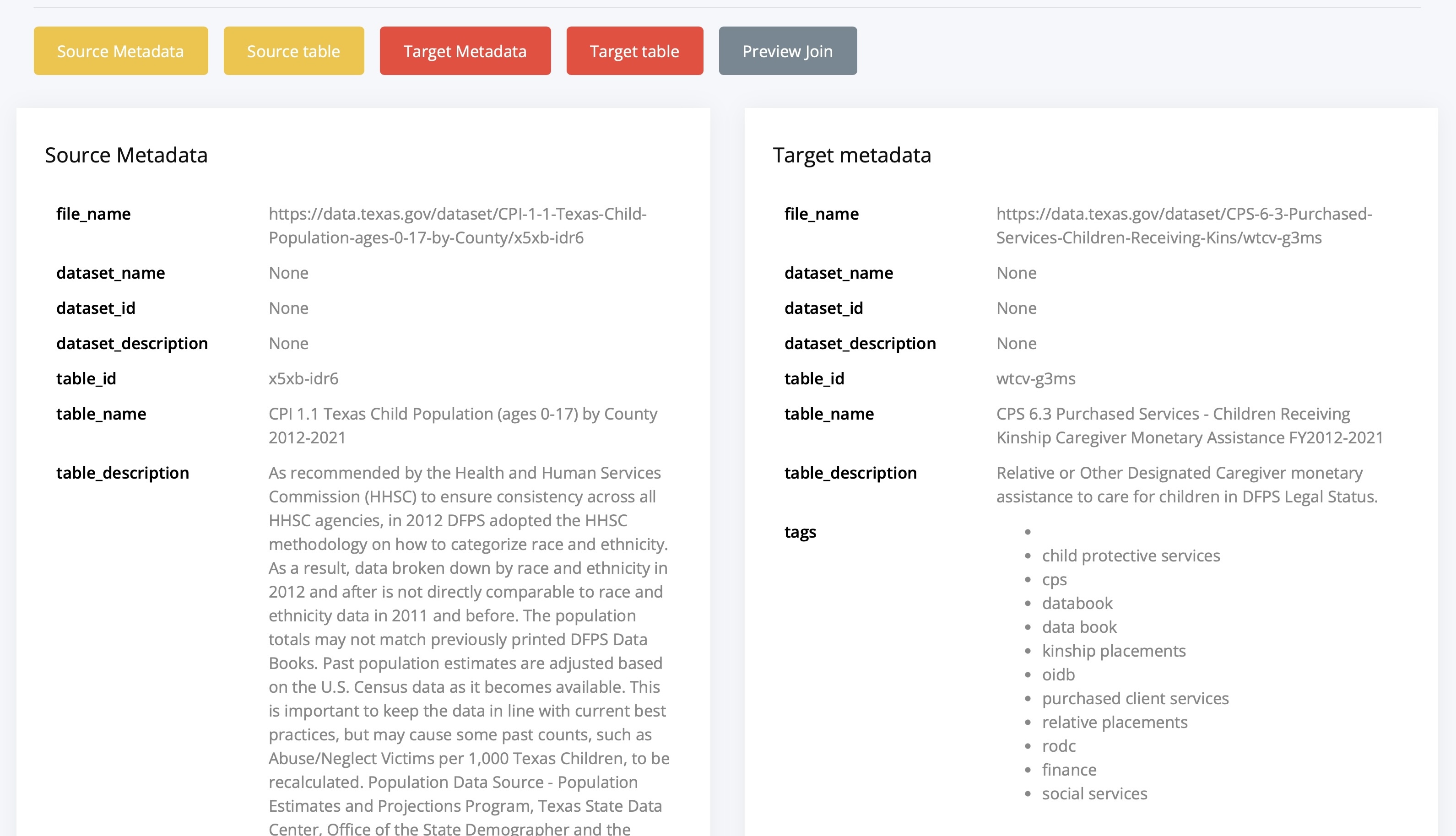}
    \caption{Snapshot of the annotation tool that was developed to collect human annotations for context-aware joinable columns in OpenData. The source metadata on the left belongs to~\cref{fig:topjoin_main}(a), the query table in~\cref{ex:context_aware_join}, and the target metadata on the right belongs to~\cref{fig:topjoin_main}(b). Annotators had access to the complete metadata and the first 1000 rows of both tables. Additionally, they were able to preview the first 1500 rows of the resulting joined table.}
    \label{fig:ReTAT}
\end{figure*}
}{}

To enable a thorough analysis of the performance of various methods for joinable column discovery, we curate a collection of datasets spanning diverse domains and characteristics. Our collection includes both academic and real-world enterprise datasets, each with distinct attributes and challenges. Moreover, each dataset employs a different strategy to establish ground truth for semantically joinable columns, tailored to its respective domain.

\introparagraph{GO Sales} 
The GO Sales database~\cite{GOSALES_browne2010ibm,GoSales_SenBSBBS20,GOSALES_Daniels} is a comprehensive collection of data related to the fictional Great Outdoors (GO) company's operations. This dataset includes information about products, inventories, sales targets, organizations, and human resources. In line with typical company databases, GO Sales features various codes that represent distinct entities, such as \verb+product_key+, \verb+country_code+, \verb+return_reason_code+, \verb+retailer_key+, etc. Given that many of these codes are numeric, their semantic distinction is minimal, and there is significant overlap between unrelated columns. This characteristic is common in most enterprise datasets, making it essential to evaluate the join search techniques for such data.
The dataset includes a collection of $151$ tables and $319$ query columns, which have been identified based on primary key foreign key relationships.

\introparagraph{CIO}
The CIO data lake is a replica of a \textbf{real-world} enterprise Operational Data Store (ODS)~\cite{inmon1995building} maintained by the Chief Information Officer's organization at IBM to support various use cases across several organizations.
This data lake houses a set of database tables that contain transactional data that are derived from a number of DB2 databases and other NoSQL sources.
The dataset includes $59$ tables, and the ground truth for joinable columns was derived by analyzing a collection of SQL query logs maintained in the ODS. The data lake features abbreviated column names and a mix of string, integer, float, and date columns. The number of rows in the tables varies from $0$ to $90$ million, with an average 

\introparagraph{Wiki Join} 
Next, to emulate the data lake scenario, we incorporate the Wiki data lake benchmark from LakeBench~\cite{LakeBench}. This data lake consists of a collection of tables generated from the Wikidata knowledge base~\cite{Wikidata_VrandecicK14}. Each table contains data about entities that belong to a certain class (e.g., Company, Person, Book, or Protein), with a label column that contains entity labels and one or more property columns containing properties of the entities (e.g., Founder, Date of Birth, Publication Year, Molecular Function). The ground truth for joinable columns in this dataset is established by using the mappings of the cell values to their corresponding entities and measuring the overlap in entities between columns. This overlap is then used as a measure of the potential for joining the columns. A majority of the columns~($57.5\%$) in this dataset are strings, resulting in a higher degree of semantics attached to each column. Moreover, since most language models are trained on Wikipedia dumps, the embeddings generated for this dataset may effectively represent the real semantics, making it suitable for assessing the semantic joinability of columns. The data lake contains $46,521$ tables and  $46,946$ query columns.

\introparagraph{OpenData}
The OpenData benchmark is a collection of tables extracted from open government data repositories, which have been frequently used in set-similarity-based join search studies~\cite{JOSIE_ZhuDNM19,LSHENSEMBLE_ZhuNPM16}. In prior work, an exact method was utilized to determine set overlap and leverage it as a benchmark for joinability evaluation. 
In our work, we gather \emph{human annotations for this dataset to enable the creation of context-aware joins that reflect real-world user behavior}.
We identified $471$ column pairs that exhibited high containment scores and solicited join labels from fifteen human annotators.
During the annotation process, human annotators had access to table snippets, metadata such as table descriptions, organization IDs, and tags, as well as the potential outcome of the joined table, as shown in Fig~\ref{fig:ReTAT}.

We received $6$ to $15$ annotations per sample and regarded columns as joinable if at least $10\%$ of the annotations were positive. 
Subsequently, we discovered only $42$ context-aware joinable column pairs among the $471$ pairs that were identified as joinable through set similarity.

\introparagraph{{Nextia$_{\boldsymbol{JD}}$} (S and M)}
Building on the work of Cong et al.~\cite{WarpGate_CongGFJD23}, we incorporate the small and medium testbeds from the Nextia Join Discovery benchmarks~\cite{Nextia_Flores}. \nextiaS is a small testbed comprising datasets with file sizes ranging from $1$ to $100$ MB, while \nextiaM is a medium testbed featuring datasets with file sizes between $100$ MB and $1$ GB. These testbeds contain tables obtained from open data repositories such as Kaggle and OpenML.\footnote{\url{https://www.essi.upc.edu/dtim/NextiaJD/}} 
The quality of joins for column pairs is assessed using a combination of the containment score and cardinality proportion. The idea is that column pairs with larger containment and a similar cardinality proportion are more likely to be joinable. Following WarpGate's approach~\cite{WarpGate_CongGFJD23}, we utilize the pairs with \verb+Good+ and \verb+High+ quality labels to establish the ground truth.

\introparagraph{{Valentine}} Koutras et al.~\cite{Valentine_KoutrasSIPBFLBK21} presented Valentine, an open-source experiment suite for automated schema matching~\cite{valentine_url}.
The experiment suites consist of 4 related table tasks (Unionable, View-Unionable, Joinable, and Semantically Joinable) of which we use \emph{semantically-joinable} benchmarks in our evaluations. 
The semantically joinable datasets consist of fabricated table pairs from 5 sources: TPC-DI, Open Data, ChEMBL, Magellan Data Repository, and Wiki Join. 
The table pairs are produced through a process that involves dividing a source table into two smaller tables by vertically and horizontally splitting it with varying degrees of column and row overlap. These divided tables are then populated with a combination of noisy/verbatim schemata (column names) and noisy instances (cell values) as the noise present in the instances may limit the performance of approaches that rely on equality joins.

\introparagraph{{AutoJoin}} Zhu et al.~\cite{AutoJoin} provides an \textbf{AutoJoin} benchmark for fuzzy joins containing joinable columns with fuzzy matching values.\footnote{\url{https://github.com/Yeye-He/Auto-Join}}
There are 31 table pairs constructed by querying Bing search and Google Web Tables. The table pairs exhibit inconsistencies such as formatting variations and the presence or absence of middle initials in names.
While the benchmark was originally designed to evaluate fuzzy joins (string matches for joinable rows), we repurpose it as a dataset for joinable column search by annotating columns that can be joined across table pairs. In this benchmark, like in Valentine dataset, the performance of approaches that rely on equality joins is expected to be limited due to fuzzy matches. 

We also attempted to use the join-search datasets presented in Deng et al. ~\cite{vldblakebench}. However, all the data lake tables mentioned in the ground truth have not been released. So, we could not include those datasets in our analysis~\footnote{The issue is raised on their GitHub: \url{https://github.com/RLGen/LakeBench/issues/4}}.

Based on the characteristics of the datasets, we group the datasets into three categories. 
The first category comprises datasets generated from \textbf{relational databases}, specifically GO Sales and CIO. Given that these tables adhere to the schema constraints, we anticipate that equality join-based methods will function effectively on these datasets.
The second category, \textbf{Data Lakes}, involves datasets that include tables originating from distinct sources, Wiki-Join, OpenData, and \nextia{}. These datasets utilized different semantic joinability concepts to establish their ground truth. Consequently, we anticipate semantic join methods to yield superior results for these datasets.
Finally, the third category consists of \textbf{Fuzzy Join} datasets, Valentine and AutoJoin, which have been intentionally crafted to evade equality join approaches. Together, these datasets provide a comprehensive testbed for evaluating joinable table discovery methods under heterogeneous real-world conditions.

\section{Methods}
\label{sec:methods}
\begin{figure*}
    \centering
    \includegraphics[scale=0.65]{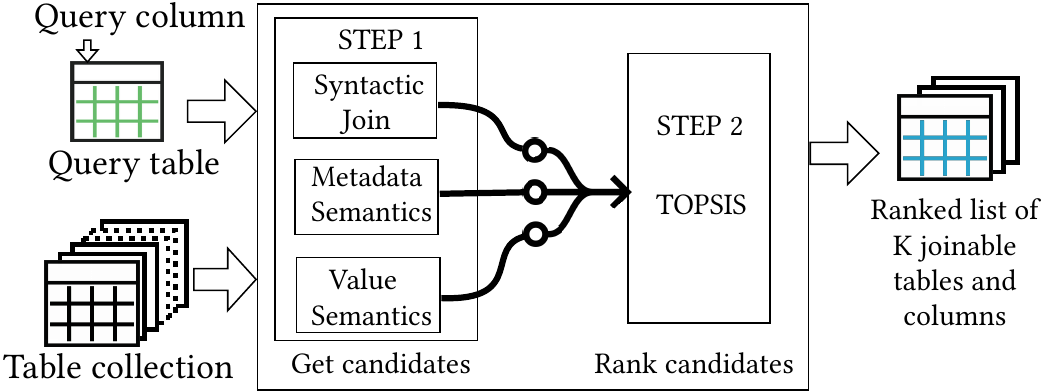}
    \caption{Architecture of the ensemble-based join discovery approach.}
    \label{fig:topjoin_architecture}
\end{figure*}

In this section, we describe the joinable table discovery methods that are included in our analysis.  We consider three broad categories of approaches—\emph{equi-join}, \emph{semantic join}, and \emph{ensemble-based}—each relying on different underlying criteria such as syntactic overlap, semantic similarity, and metadata context.  The goal of our analysis is not to introduce a new algorithm or to compare all the existing methods, but to study how these criteria, individually and in combination, influence the effectiveness of context-aware join discovery.

\subsection{Equi-join Methods}
Zhu et al.~\cite{LSHENSEMBLE_ZhuNPM16} introduced \textbf{LSH Ensemble}, where two columns are defined as joinable if their set containment surpasses a specified threshold. 
Set containment is computed as the ratio of unique overlapping values between the query and data lake column to the total unique values in the query. 
To identify columns in data lake tables with substantial set containment, they devised a Locality-Sensitive Hashing (LSH) Ensemble index designed to approximate the computation of set containment. 
Zhu et al.~\cite{JOSIE_ZhuDNM19} proposed \textbf{JOSIE}, which also searches for columns in a data lake that can be joined with the query column. 
JOSIE specifically considers an overlap only when the values in two columns are identical. 
Unlike threshold-based LSH Ensemble, JOSIE is a top-$k$ search method that relies on an inverted index and conducts searches for columns using an exact set containment criterion.
Furthermore, Esmailoghli et al.~\cite{MATE_EsmailoghliQA22} presented \textbf{MATE}, which extends JOSIE—allowing users to choose multiple columns collectively as a query to search for joinable data lake tables.

\subsection{Fuzzy Join Methods}
While equi-join operations rely on exact value matches between columns, fuzzy string similarity metrics have also been proposed in the literature~\cite{AutoJoin,WangLF11}. 
Fuzzy joins consider not only exact matches but also token matching, character matching, or generalized edit distance~\cite{AutoJoin,Valentine_KoutrasSIPBFLBK21}. 
In Auto-Join, Li et al.~\cite{2021_li_autojoin} proposed a transformation program that fuzzily joins two tables by maximizing matching recall for a given precision target. 
SEMA-JOIN~\cite{SEMAJOIN_HeGC15} takes a data-driven approach to learn statistical correlations between cell values to find join pairs. 
MF-Join~\cite{MFJoin_WangLZ19} proposed a multi-level filtering approach for fuzzy string similarity join. 
These approaches are highly efficient at identifying which pairs of rows to join once two joinable tables are known. 
However, they do not scale to the broader problem of searching for joinable tables in large data lakes and are therefore not included in our analysis.

\subsection{Semantic Join Methods}
Fuzzy join methods, despite handling string transformations, may still miss relationships such as abbreviations or synonyms. 
To address such cases, several \emph{semantic join} methods have been proposed. 
Dong et al.~\cite{PEXESO_DongT0O21} introduced \textbf{PEXESO}, which incorporates synonym and abbreviation matching when searching for top-$k$ data lake columns that are semantically joinable with a query column. 
Subsequently, \textbf{DeepJoin}~\cite{DeepJoin_Dong0NEO23} improved the effectiveness and efficiency of PEXESO by fine-tuning a pretrained language model (PLM) for this task. 
Columns are transformed into text and then embedded using various PLMs, including FastText~\cite{FastText_MikolovGBPJ18}, BERT~\cite{BERT_DevlinCLT19}, and MPNet~\cite{MPNet_Song0QLL20}, as well as fine-tuned variants such as DistilBERT and MPNet. 
The fine-tuned model embeds each column into a vector space such that joinable columns are close to each other, and searches are accelerated using the Hierarchical Navigable Small World (HNSW) index~\cite{HNSW_MalkovY20}.

\textbf{WarpGate}~\cite{WarpGate_CongGFJD23} is another recent semantic join method that also determines joinability based on embeddings of column values. 
It embeds each data lake column into a high-dimensional embedding space. 
Unlike DeepJoin, which fine-tunes a PLM, WarpGate uses pre-trained WebTable embeddings (FastText embeddings pre-trained for web tables~\cite{fasttext_tabular_TGL21})\footnote{\url{https://huggingface.co/ddrg/web_table_embeddings_combo64}} to represent columns. 
When a user provides a query column, it is embedded in the same way, and the top-$k$ most similar columns are retrieved via SimHash LSH. 

Other related work on table discovery includes unionable table search~\cite{TUS_NargesianZPM18,SANTOS_KhatiwadaFSCGMR23,D3L,Aurun_FernandezAKYMS18,DBLP:conf/sigmod/KhatiwadaSM23}. 
Unlike join search, union search techniques focus on appending new rows to the query table rather than additional columns, expanding the table vertically.

\subsection{Ensemble-Based Approach for Context-Aware Joins}

Since context-aware join discovery depends on multiple complementary factors, the optimal approach is often one that integrates several joinability criteria rather than relying on a single measure. 
To analyze this effect, we adopt an \emph{ensemble-based framework} that aggregates syntactic, semantic, and contextual signals through multi-criteria ranking.\footnote{Preliminary results of this approach were presented as \emph{TOPJoin: A Context-Aware Multi-Criteria Approach for Joinable Column Search} at the TaDA 2025 Workshop co-located with VLDB~\cite{KokelEtAl2025TOPJoin}.}
Its overall architecture is illustrated in~\cref{fig:topjoin_architecture}.

Given a query table with a designated join column, the approach first identifies potential join candidates from a collection of tables (e.g., a data lake) using three complementary strategies:  
(1) \textbf{Syntactic Join}, which measures the overlap in normalized cell values;  
(2) \textbf{Metadata Semantics}, which measures semantic similarity of table and column metadata; and  
(3) \textbf{Value Semantics}, which measures semantic similarity of the column values themselves.  
Implementation details of these components are discussed in~\cref{sec:experiments}. 

After candidate generation, the approach scores and re-ranks these candidates using the six joinability criteria described in~\cref{sec:problem_definition}.  
The individual criterion scores are then combined using the \textbf{TOPSIS} (Technique for Order Preference by Similarity to Ideal Solution) algorithm~\cite{TOPSIS_HwangY81}, which provides a principled framework for multi-criteria optimization.  
TOPSIS ranks candidates according to their distance from an ideal solution—preferring those closest to the positive ideal and farthest from the negative one.  
The procedure begins with a normalized decision matrix, where each row corresponds to a candidate and each column to a criterion.  
If the criteria have different relative importance, the matrix can be reweighted accordingly.  
The distance of each candidate from the ideal solution is then computed, yielding a composite ranking that balances the contributions of all criteria.  

This ensemble formulation allows us to systematically study how combining multiple dimensions of joinability—syntactic overlap, semantic similarity, and metadata context—affects discovery effectiveness across diverse datasets.

\section{Experiments}
\label{sec:experiments}

In this section, we present a detailed empirical analysis of joinable column discovery approaches across diverse datasets and scenarios. 
Our goal is to characterize how different techniques and joinability criteria behave under varying data characteristics, rather than to optimize any single method. We organize our study around three research questions:
\begin{enumerate}
\item How do different equi-join, semantic join and ensemble approaches perform for the context-aware join search task?\label{Q1:comparemethods}
\item How do the six criteria influence the effectiveness of context-aware join discovery?\label{Q2:comparecriteria}
\item In absence of finetuning data, which pre-trained embeddings are most useful for the context-aware join?\label{Q3:compareembeddings} 
\end{enumerate}

\subsection{Implementation Details}

\introparagraph{Equi-join Method} For our experimental analysis, we use LSH Ensemble~\cite{LSHENSEMBLE_ZhuNPM16} as an equi-join approach. We use its public implementation with default parameters.\footnote{\url{https://github.com/ekzhu/datasketch}} We also include a top-k exact set containment as a search baseline (\textbf{Exact Set Intersection}) to understand if naive set containment already gives better joinability effectiveness than complex methods. Specifically, we follow equi-join-based JOSIE~\cite{JOSIE_ZhuDNM19} to create this baseline. As we only compare this method's effectiveness against other methods, our implementation generates the same result as JOSIE, but we do not utilize all the runtime optimizations suggested in the original paper. 

\introparagraph{Semantic Join Methods} Furthermore, we use DeepJoin~\cite{DeepJoin_Dong0NEO23} as one of the semantic join search methods. Due to the unavailability of finetuned model and the training data, we have replicated DeepJoin model based on the publicly available implementation\footnote{\url{https://github.com/RLGen/LakeBench/tree/main/join/Deepjoin}} by training it on OpenData. WarpGate~\cite{WarpGate_CongGFJD23} is another recent semantic join method that we include in the experiments. Since its code is not available publicly, we reproduce WarpGate using the information given in the paper. As WarpGate is based on FastText WebTable Embeddings, for a comprehensive analysis, we implement its three different variations. Specifically, we implement WarpGate using two openly available variations of WebTable-based FastText Embeddings~\cite{fasttext_tabular_TGL21}. Next, we implement it using original FastText embeddings~\cite{FastText_MikolovGBPJ18} and use the best-performing embeddings among three for comparison.
WarpGate is shown to outperform Aurum~\cite{Aurun_FernandezAKYMS18} and D3L~\cite{D3L} for semantic join search tasks.

\introparagraph{Ensemble Methods} We next describe how we implement the \ourmethod{} ensemble-based method for multi-criteria join discovery used in our analysis, including its three candidate-generation strategies and the criteria scoring process.
The first strategy utilizes \emph{syntactic joinability measures}, which have been widely adopted by various approaches~\cite{JOSIE_ZhuDNM19,LSHENSEMBLE_ZhuNPM16,LSHForest_BawaCG05}. Essentially, we want to search for the columns that have maximum overlapping values. For this, we follow JOSIE~\cite{JOSIE_ZhuDNM19} method and build a posting list for each normalized cell value which points to a list of columns that contains that value. Then we store the posting list in the form of an inverted index.
During search time, we use the inverted index to find the columns with the highest value overlap. 

Note that the syntactic join candidate identification strategy within \ourmethod{} maintains an inverted index for cell values. The size of this inverted index can grow quickly for large data lakes. In our experiments, we limit the size by sampling rows from the tables. However, this might still not be sufficient. So, we examine the utility of an approximate approach using MinHashes. As MinHashes can efficiently estimate Jaccard similarity, we compute MinHashes for each column (with $100$ permutations) and create an index using the Hamming distance. At search time, we query the index for $k$ closest columns and use that as syntactic join candidates. We continue to utilize samples of original column values to compute the MinHashes. However, we maintain only the MinHash index instead of the inverted index. This choice leads to significant reductions in memory requirements. For instance, the inverted index for the CIO column, which had a size of $2.6$GB, is replaced by a MinHash index that is only $508$KB in size. This reduction in memory usage is particularly beneficial for handling large datasets. We refer to this approximate version as \textbf{\ourmethod-MinHash}.

For the second strategy, we utilize a \emph{metadata semantics} technique to identify potential table-column join candidates that are semantically related to the query table and column. Our method entails crafting a metadata sentence for each column by combining various metadata elements such as the table name, table description, column details, dataset source, the query column's name and description, etc. Subsequently, employing a sentence transformer~\cite{sbert}, we produce sentence embeddings for each column and establish an index. When conducting searches, we employ K-nearest neighbor search within the embedding space to identify candidate columns sharing similar contextual attributes with the query's metadata.

Lastly, for the third strategy, \emph{value semantics}, we transform each data lake column into text by concatenating its values, and embed it using a sentence transformer. We similarly embed the query column and search for data lake columns (join candidates) having embeddings that are closest to the query column.

\subsection{Results}

\begin{table*}[!ht]
\footnotesize
\caption{Comparison of \ourmethod{} with LSH Ensemble, DeepJoin, and WarpGate for all the datasets (\cref{sec:datasets}); at K=10. Best results are indicated in bold and second best are underlined.}
\label{tab:topjoin_results}
\centering
\renewcommand{\arraystretch}{0.65}
\resizebox{1\linewidth}{!}{
\begin{tabular}{c|c||c|c||c|c|c||c|c||c}
\toprule
\multirow{2}{*}{\textbf{Method}} &  \textbf{Metric} & \multicolumn{2}{c||}{\textbf{Relational DBs}}  & \multicolumn{3}{c||}{\textbf{Data Lakes}} & \multicolumn{2}{c||}{\textbf{Fuzzy Joins}} & \multirow{2}{*}{\textbf{Avg.}} \\   
   & \textbf{@ K=10}  & \textbf{GO Sales}  & \textbf{CIO} &  \textbf{Wiki-Join} & \textbf{OpenData} & \textbf{Nextia$_{\boldsymbol{JD}}$} & \textbf{Valentine} &  \textbf{AutoJoin} & \\ 

\midrule
\multirow{3}{*}{LSH Ensemble} &MRR &0.31 &0.29 & 0.80 &0.44 &0.34 &0.34 &0.21 &0.39 \\
 &MAP &0.29 &0.25 & 0.34 &0.37 &0.34 &0.34 &0.21 &0.30 \\
 &Recall &0.62 &0.41 & 0.37 &0.46 &0.69 &0.45 &0.33 &0.48 \\
\midrule
\multirow{3}{*}{DeepJoin} & MRR & 0.29 & 0.20 & 0.68 & 0.44 & 0.27 & 0.66  & \best{0.41} &0.42\\
 &MAP & 0.26 & 0.16 & 0.26 & 0.36 & 0.27 & 0.66 &  \best{0.41} & 0.34\\
 &Recall & 0.54 & 0.29 &  0.35 & 0.47 & 0.39 & 0.88 &  \best{0.80} & 0.53\\
\midrule
\multirow{3}{*}{WarpGate} &MRR & \best{0.68} & 0.33 &0.67 & 0.58 &0.28 &0.58 &0.24 &0.47 \\
 &MAP & \best{0.66} & 0.27 &0.21 & 0.46 &0.28 &0.58 &0.24 &0.38 \\
 &Recall & \best{0.95} & 0.46 &0.26 & \secondbest{0.63} &0.53 &0.83 &0.5 &0.57 \\
 \midrule
\multirow{3}{*}{\ourmethod{}} &MRR & \secondbest{0.52} &\best{0.39} & \best{0.91} & \best{0.61} & \best{0.40} & \best{0.81} & \secondbest{0.32} & \best{0.56} \\
 &MAP & \secondbest{0.52} & \best{0.34} & \best{0.50} & \best{0.49} & \best{0.40} & \best{0.81} & \secondbest{0.32} & \best{0.48} \\
&Recall & \secondbest{0.94} & \best{0.68} & \best{0.53} & \best{0.65} & \best{0.65} &\best{0.96} & \secondbest{0.60} & \best{0.71} \\
 \midrule
 \multirow{3}{*}{\ourmethod{}-MinHash} &MRR &\secondbest{0.52} &\secondbest{0.37} &\secondbest{0.90} & \secondbest{0.60} & \textbf{0.40} & \textbf{0.81} &{0.28} & \best{0.56} \\
 &MAP &\secondbest{0.52} &\secondbest{0.33} &\secondbest{0.49} & \secondbest{0.48} & \textbf{0.40} & \textbf{0.81} &{0.28} &\secondbest{0.47} \\
&Recall &\secondbest{0.94} & \secondbest{0.60} & \best{0.53} & 0.62 & \textbf{0.65} & \textbf{0.96} & {0.50} &\secondbest{0.69} \\
\midrule
\midrule
\textcolor{gray}{Exact} & \textcolor{gray}{MRR} & \textcolor{gray}{0.42} & \textcolor{gray}{0.45} & \textcolor{gray}{0.90} & \textcolor{gray}{0.54} & \textcolor{gray}{0.45} & \textcolor{gray}{0.55} & \textcolor{gray}{0.23}& \textcolor{gray}{0.51} \\ 
\textcolor{gray}{Set} & \textcolor{gray}{MAP} & \textcolor{gray}{0.40} & \textcolor{gray}{0.40} & \textcolor{gray}{0.47} & \textcolor{gray}{0.48} & \textcolor{gray}{0.45} & \textcolor{gray}{0.55} & \textcolor{gray}{0.23}& \textcolor{gray}{0.43} \\ 
 \textcolor{gray}{Intersection} & \textcolor{gray}{Recall} & \textcolor{gray}{0.74} & \textcolor{gray}{0.71} & \textcolor{gray}{0.51} & \textcolor{gray}{0.59} & \textcolor{gray}{0.81} & \textcolor{gray}{0.66} & \textcolor{gray}{0.33}& \textcolor{gray}{0.62} \\ 
\bottomrule
\end{tabular}
} 
\end{table*}

\begin{figure*}[!h]
\centering
   \begin{subfigure}[b]{0.9\textwidth}
        \inputtikz{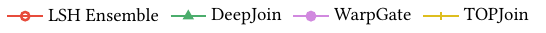}
    \end{subfigure}
   \resizebox{0.9\textwidth}{!}{    {

 \begin{subfigure}[b]{0.30\textwidth}
    \inputtikz{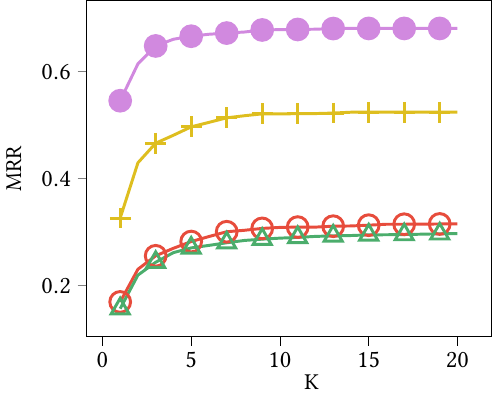}
    \vspace{-1\baselineskip}
     \caption{MRR on GO Sales}
    \label{fig:mrr_go_sales}
    \end{subfigure}
\hfill
\begin{subfigure}[b]{0.30\textwidth}
    \inputtikz{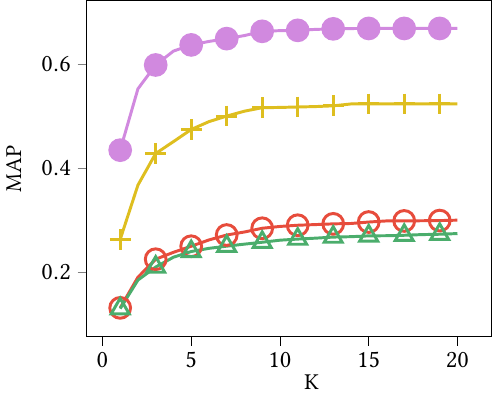}
     \vspace{-1\baselineskip}
     \caption{MAP on GO Sales}
    \label{fig:map_go_sales}
    \end{subfigure}
\hfill
\begin{subfigure}[b]{0.30\textwidth}
    \inputtikz{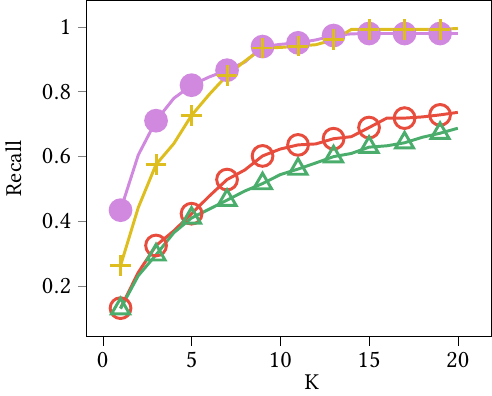}
    \vspace{-1\baselineskip}
     \caption{Recall on GO Sales}
    \label{fig:recall_go_sales}
    \end{subfigure}
}}
    \vspace{\baselineskip}
   \resizebox{0.9\textwidth}{!}{{
 \begin{subfigure}[b]{0.30\textwidth}
\inputtikz{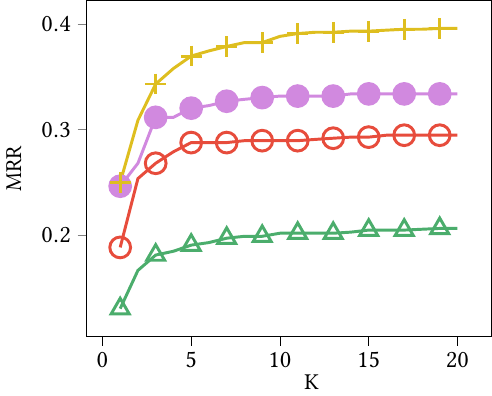}
    \vspace{-1\baselineskip}
     \caption{MRR on CIO}
    \label{fig:mrr_cio}
    \end{subfigure}
\hfill
\begin{subfigure}[b]{0.30\textwidth}
\inputtikz{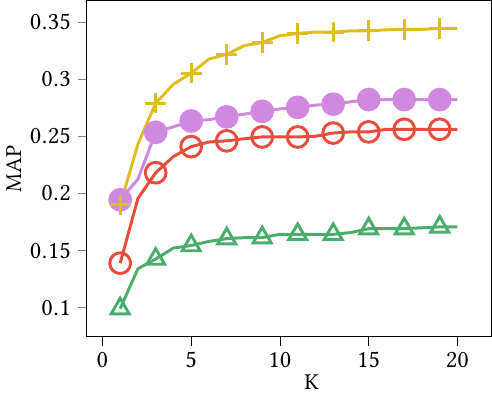}
    \vspace{-1\baselineskip}
     \caption{MAP on CIO}
    \label{fig:map_cio}
    \end{subfigure}
\hfill
\begin{subfigure}[b]{0.30\textwidth}
\inputtikz{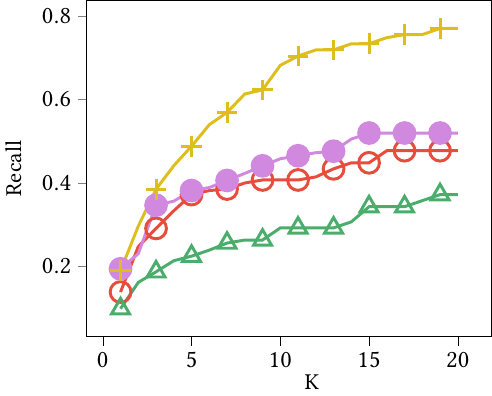}
    \vspace{-1\baselineskip}
     \caption{Recall on CIO}
    \label{fig:recall_cio}
    \end{subfigure}

}}
\caption{Effectiveness of different methods in Relational DB Benchmarks.}
\label{fig:effectiveness_relational_dbs}
\end{figure*}

\begin{figure*}[htbp!]
    \vspace{\baselineskip}
\centering
   \begin{subfigure}[b]{0.9\textwidth}
        \inputtikz{effectiveness_legend}
    \end{subfigure}
    \vspace{\baselineskip}
    
\resizebox{0.9\textwidth}{!}{{
    \centering
 \begin{subfigure}[b]{0.30\textwidth}
    \inputtikz{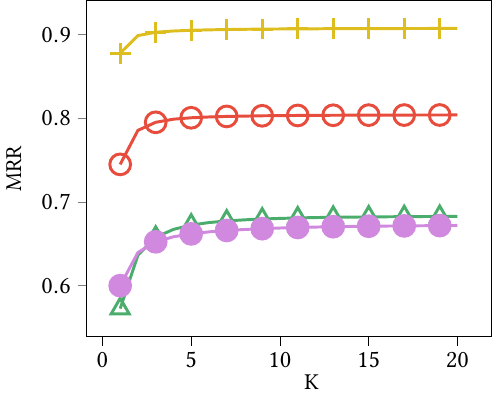}
    \vspace{-1\baselineskip}
     \caption{MRR on Wiki-Join}
    \label{fig:mrr_wiki_join}
    \end{subfigure}
\hfill
\begin{subfigure}[b]{0.30\textwidth}
    \inputtikz{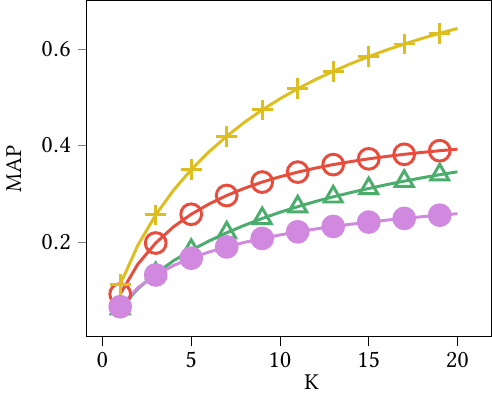}
    \vspace{-1\baselineskip}
     \caption{MAP on Wiki-Join}
    \label{fig:map_wiki_join}
    \end{subfigure}
\hfill
\begin{subfigure}[b]{0.30\textwidth}
    \inputtikz{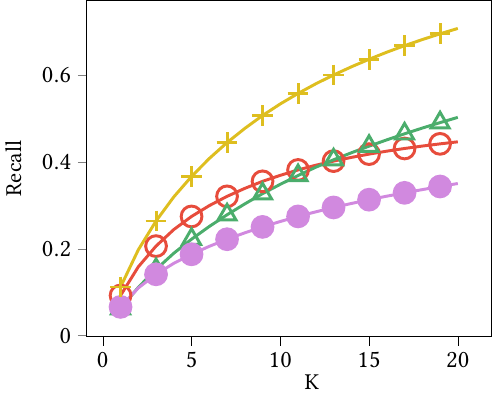}
    \vspace{-1\baselineskip}
     \caption{Recall on Wiki-Join}
    \label{fig:recall_wiki_join}
    \end{subfigure}
}}
    \vspace{\baselineskip}

\resizebox{0.9\textwidth}{!}{{
\begin{subfigure}[b]{0.30\textwidth}
    \inputtikz{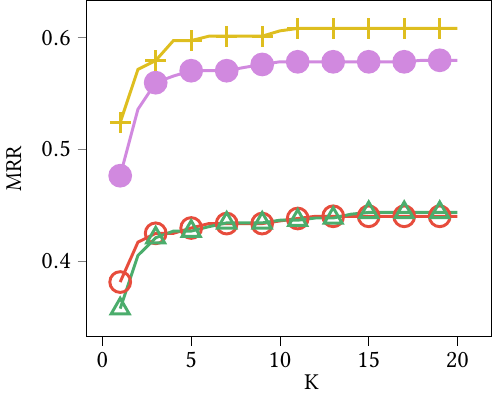}
    \vspace{-1\baselineskip}
     \caption{MRR on OpenData}
    \label{fig:mrr_open_data}
    \end{subfigure}
\hfill
\begin{subfigure}[b]{0.30\textwidth}
    \inputtikz{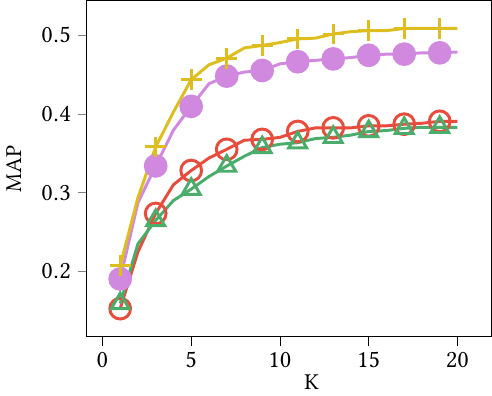}
    \vspace{-1\baselineskip}
     \caption{MAP on OpenData}
    \label{fig:map_open_data}
    \end{subfigure}
\hfill
\begin{subfigure}[b]{0.30\textwidth}
    \inputtikz{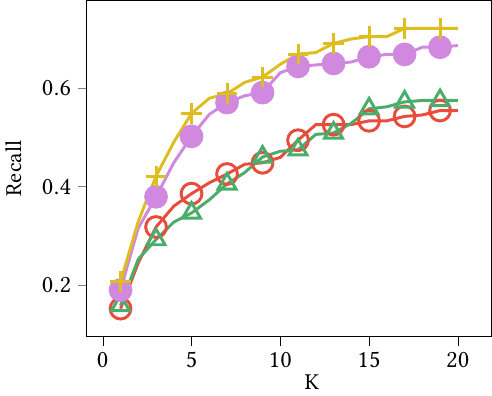}
    \vspace{-1\baselineskip}
     \caption{Recall on OpenData}
    \label{fig:recall_open_data}
    \end{subfigure}
}}
    \vspace{\baselineskip}

\resizebox{0.9\textwidth}{!}{{
\begin{subfigure}[b]{0.30\textwidth} \inputtikz{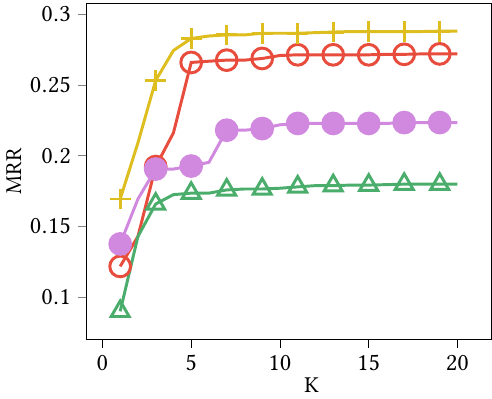}
    \vspace{-1\baselineskip}
     \caption{MRR on Nextia$_{\mathbf{JD}}$-M}
    \label{fig:mrr_nextiajd-M}
    \end{subfigure}
\hfill
\begin{subfigure}[b]{0.30\textwidth}
    \inputtikz{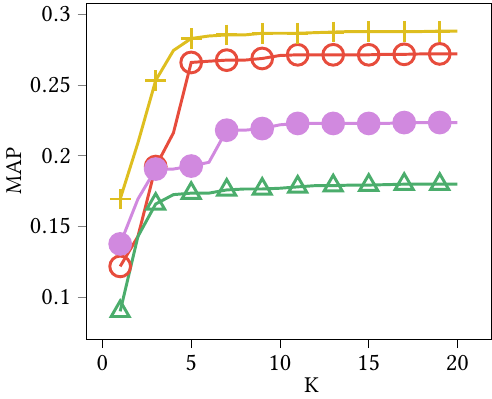}
    \vspace{-1\baselineskip}
     \caption{MAP on Nextia$_{\mathbf{JD}}$-M}
    \label{fig:map_nextiajd-M}
    \end{subfigure}
\hfill
\begin{subfigure}[b]{0.30\textwidth}
    \inputtikz{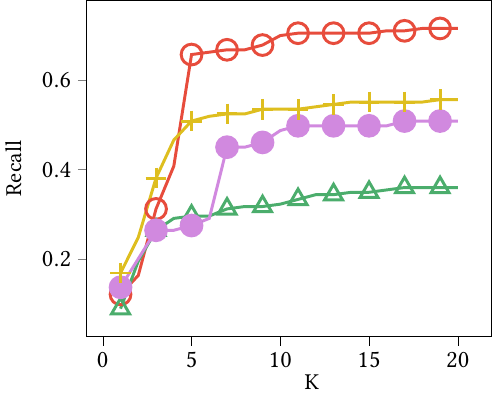}
    \vspace{-1\baselineskip}
     \caption{Recall on Nextia$_{\mathbf{JD}}$-M}
    \label{fig:recall_nextiajd-M}
    \end{subfigure}
}}

    \vspace{\baselineskip}
    
\resizebox{0.9\textwidth}{!}{{

\begin{subfigure}[b]{0.30\textwidth} \inputtikz{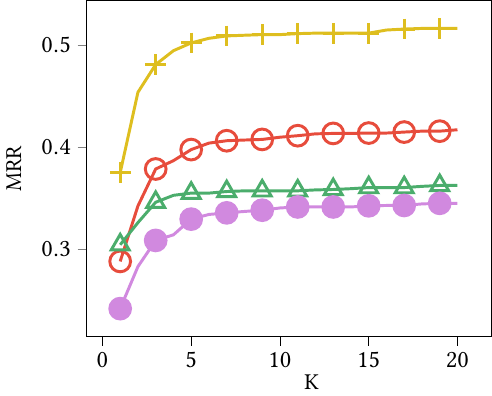}
    \vspace{-1\baselineskip}
     \caption{MRR on Nextia$_{\mathbf{JD}}$-S}
    \label{fig:mrr_nextiajd-S}
    \end{subfigure}
\hfill
\begin{subfigure}[b]{0.30\textwidth}
    \inputtikz{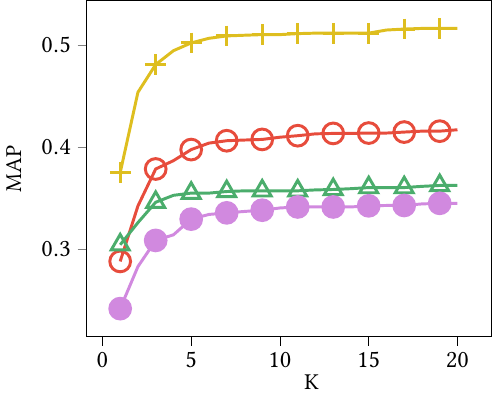}
    \vspace{-1\baselineskip}
     \caption{MAP on Nextia$_{\mathbf{JD}}$-S}
    \label{fig:map_nextiajd-S}
    \end{subfigure}
\hfill
\begin{subfigure}[b]{0.30\textwidth}
    \inputtikz{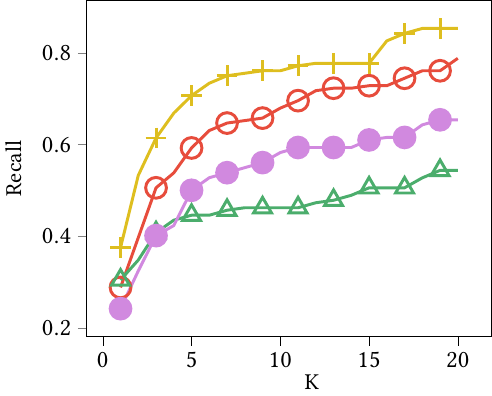}
    \vspace{-1\baselineskip}
     \caption{Recall on Nextia$_{\mathbf{JD}}$-S}
    \label{fig:recall_nextiajd-S}
    \end{subfigure}

}
}
    \vspace{\baselineskip}
\caption{Effectiveness of different methods in Data lake Benchmarks.}
\label{fig:effectiveness_data_lakes}
\end{figure*}

\begin{figure*}[!ht]
   \begin{subfigure}[b]{0.9\textwidth}
        \inputtikz{effectiveness_legend}
    \end{subfigure}
        \centering
   \resizebox{0.9\textwidth}{!}{ {

 \begin{subfigure}[b]{0.30\textwidth}
\inputtikz{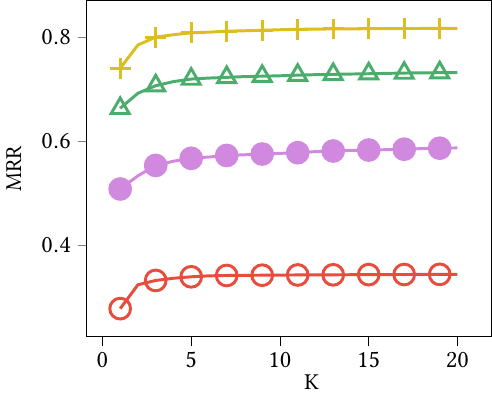}
    \vspace{-1\baselineskip}
     \caption{MRR on Valentine}
    \label{fig:mrr_valentine}
    \end{subfigure}
\hfill
\begin{subfigure}[b]{0.30\textwidth}
\inputtikz{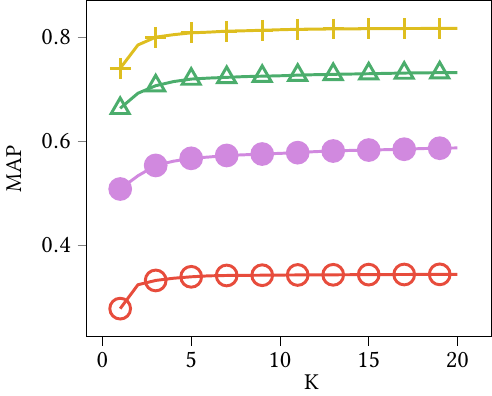}
    \vspace{-1\baselineskip}
     \caption{MAP on Valentine}
    \label{fig:map_valentine}
    \end{subfigure}
\hfill
\begin{subfigure}[b]{0.30\textwidth}
\inputtikz{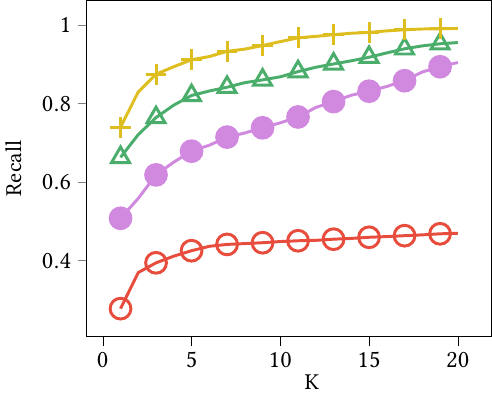}
    \vspace{-1\baselineskip}
     \caption{Recall on Valentine}
    \label{fig:recall_valentine}
    \end{subfigure}
    }}

        \vspace{\baselineskip}

   \resizebox{0.9\textwidth}{!}{{
 \begin{subfigure}[b]{0.30\textwidth}
    \inputtikz{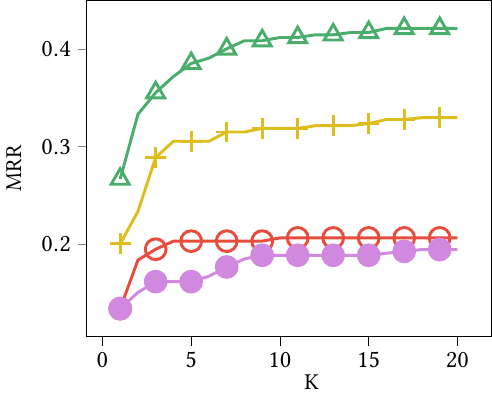}
    \vspace{-1\baselineskip}
     \caption{MRR on AutoJoin}
    \label{fig:mrr_autojoin}
    \end{subfigure}
\hfill
\begin{subfigure}[b]{0.30\textwidth}
    \inputtikz{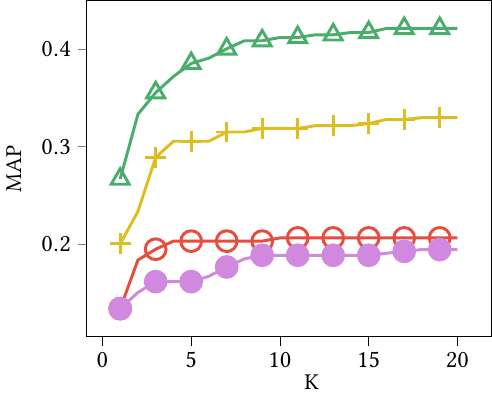}
     \vspace{-1\baselineskip}
     \caption{MAP on AutoJoin}
    \label{fig:map_autojoin}
    \end{subfigure}
\hfill
\begin{subfigure}[b]{0.30\textwidth}
    \inputtikz{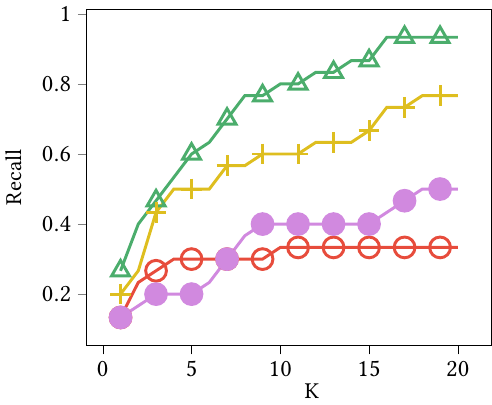}
    \vspace{-1\baselineskip}
     \caption{Recall on AutoJoin}
    \label{fig:recall_autojoin}
    \end{subfigure}

}}

\caption{Effectiveness of different methods in Fuzzy Join Benchmarks.}
\label{fig:effectiveness_fuzzy_joins}
\end{figure*}

We now report the context-aware join discovery effectiveness of one equi-join approach (\textbf{LSH Ensemble}), two semantic join approaches (\textbf{DeepJoin} and \textbf{WarpGate}), and the ensemble-based \textbf{TOPJoin} approach across the seven datasets described in~\cref{sec:datasets}.\footnote{The code and artifacts required to replicate the results are available at \url{https://ibm.biz/context-aware-join}.}
As our focus in this paper is on accuracy, we use standard effectiveness metrics: Mean Reciprocal Rank (MRR), Mean Average Precision (MAP), and Recall@K. 
Results for \(K = 10\) are reported in~\cref{tab:topjoin_results}, and trends for MRR, MAP, and Recall across the top 1--20 results are shown in~\cref{fig:effectiveness_relational_dbs,fig:effectiveness_data_lakes,fig:effectiveness_fuzzy_joins}. 
Note that the Exact Set Intersection method requires maintaining an index for every cell value in millions of tables and columns, which does not satisfy the compute and storage constraints of enterprise-scale systems. 
Therefore, we include these results only for completeness; they are shown separately in gray at the bottom of~\cref{tab:topjoin_results}.

For \textbf{relational DBs}, \ourmethod{} and WarpGate are among the top performants. The equality join approaches, which we anticipated to be effective on these datasets, failed for GO Sales as they encountered a high intersection between distinct integer columns, such as \texttt{product\_key} and \texttt{retainer\_key}. Although the exact approach performed the best on the CIO dataset, it is infeasible in practice, since the column size of that data set reaches up to $90$ million.

For \textbf{Data Lakes}, \ourmethod{} demonstrates superior performance compared to all the approaches. Moreover, there is no distinct second-place finisher among the competing approaches. Notice that even though the joinable column ground truth in Wiki-Join was established using a semantic joinability notion, i.e. by mapping cell values to their corresponding entities and finding overlap in entities, LSH Ensemble performs reasonably well. This is because the dataset has a high value overlap among semantically joinable columns and low overlap among semantically non-joinable columns as indicated by the results with exact set intersection results. On the other two datasets, however, we see that equi-join approaches do not perform well. As previewed in the~\cref{ex:context_aware_join}, OpenData has a high overlap of values among semantically non-joinable columns which explains the significant difference in the performance.

\ourmethod{} excels in \textbf{Fuzzy Join domains}, demonstrating superior performance among all approaches. DeepJoin has a substantial performance difference from the other semantic join approach, WarpGate. 
WarpGate and DeepJoin share the same FastText encoder to generate column embeddings in our implementation. So, the primary distinction in their performance can be traced back to the additional context used by DeepJoin, which incorporates table names, column names, and column statistics. This extra context, absent in WarpGate, significantly contributes to the performance gap between the two approaches. This shows the importance of accounting for context in searching for joinable columns.

Across all the datasets, on average, we see that \ourmethod{} and \ourmethod-MinHash perform the best (see the last column in~\cref{tab:topjoin_results}). While using \ourmethod{}'s approximate version (\ourmethod-MinHash), there is negligible impact on the MAP and Recall values with a significant reduction in the storage requirements, which is crucial for enterprise products. This answers our research question (\ref{Q1:comparemethods}), in that while the equi-join and semantic-join approaches are useful, an ensemble approach that optimizes for multiple-criteria is more reliable.

\subsection{Key Criteria}

\begin{table}[t!]
    \caption{Comparison of different criterion on OpenData}
    \vspace{-5pt}
    \label{table:opendata_ranker_only}
    \centering
\begin{tabular}{|l|c|c|c|}
\toprule
\textbf{ Criterion} & \textbf{MRR} & \textbf{MAP} & \textbf{Recall} \\
\midrule
UNIQUE VALUES & 0.06 & 0.02 & 0.34 \\
INTERSECTION SIZE & 0.57 & 0.50 & 0.75 \\
JOIN SIZE & 0.50 & 0.43 & 0.72 \\
REVERSE JOIN SIZE & 0.47 & 0.41 & 0.75 \\
VALUE SEMANTICS & {0.75} & {0.65} & {0.82} \\
DISJOINT VALUE SEMANTICS & 0.59 & 0.38 & 0.62 \\
METADATA SEMANTICS & 0.56 & 0.42 & 0.66 \\
\midrule
\textcolor{gray}{ALL PREFERENCES} & \textcolor{gray}{0.61} & \textcolor{gray}{0.49} & \textcolor{gray}{0.65} \\
\bottomrule
\end{tabular}
\vspace{-12pt}
\end{table}

\begin{table}[t!]
\caption{Comparison of different criterion on AutoJoin}
\vspace{-5pt}
\label{table:autojoin_ranker_only}
    \centering
\begin{tabular}{|l|c|c|c|}
\toprule
\textbf{Criterion } & \textbf{MRR} & \textbf{MAP}  & \textbf{Recall}  \\
\midrule
UNIQUE VALUES & 0.04  & 0.04 & 0.66  \\
INTERSECTION SIZE &  0.24 & 0.24 & 0.7 \\
JOIN SIZE & 0.24  & 0.24 & 0.66  \\
REVERSE JOIN SIZE & 0.24 & 0.24 & 0.66 \\
VALUE SEMANTICS & 0.27 &  0.27 & 1.0 \\
DISJOINT VALUE SEMANTICS & 0.21 & 0.21 & 1.0 \\
METADATA SEMANTICS & 0.22 &  0.22 & 0.8  \\
\midrule
\textcolor{gray}{ALL PREFERENCES} & \textcolor{gray}{0.32} &  \textcolor{gray}{0.32} &  \textcolor{gray}{0.6}   \\
\bottomrule
\end{tabular}
\vspace{-12pt}
\end{table}

Next we examine six \ourmethod{} criteria and their role in facilitating context-aware join discovery. Here we consider the join-size and reverse join-size as separate criteria and hence, consider seven criteria. We report results on OpenData benchmark that has significant amount of metadata and fuzzy-match based AutoJoin benchmark, where exact string matches are rare.

To evaluate the importance of each criteria we use the TOPJoin approach where we use three approaches of identifying candidate columns and then rank these candidates using only the mentioned single criterion.
Table~\ref{table:opendata_ranker_only} and Table~\ref{table:autojoin_ranker_only} compares the performance of TOPJoin with different ranking criterion.  Based on the results, \textbf{the value semantics criterion appears to be the best criterion to rank the candidates}. Given the semantic nature of both the datasets, the fact that Value Semantics criterion outperforms other criterion is expected. 
In fact, in OpenData (\cref{table:opendata_ranker_only}), using Value Semantics criterion alone outperforms the search when all the criteria are considered (shown in the last row). This might be explained by the fact tables in OpenData are not relational and do not really contain primary keys. So the unique values and size criteria might not be extremely helpful here. The second best performer is the disjoint value semantics criterion which shows that the data lakes can have significant overlap of values within the non-joining columns and hence, it is important to consider the semantics of non-overlapping values between the join columns when making a joinability decision.

We also drop one criterion from TOPJoin to see the impact of that criterion on context-aware join search task. The results are presented in~\cref{table:opendata_without_ranker} and~\cref{table:autojoin_without_ranker}.
Based on the results on OpenData, where we remove a criterion from the preferences (\cref{table:opendata_without_ranker}), it is clear that unique-values and three size criteria (intersection size, join size and reverse join size) have minimal impact on performance.
\textbf{Removing  metadata semantics criterion has the most significant impact.} To understand why, consider the metadata such as column names and table names for \cref{fig:topjoin_main}(a) and (b) from~\cref{ex:context_aware_join} 
which are also shown in~\cref{fig:ReTAT}. 
Notice that neither the column name nor the table name indicate that \cref{fig:topjoin_main}(b) is related to Texas. However, the filename (source URL of the file) indicates that the table is related to Texas. Without this metadata, it becomes significantly harder for embeddings to identify relevant tables.

\begin{table}[]
\caption{Performance comparison when the mentioned criterion is removed from the preferences in OpenData }
\label{table:opendata_without_ranker}
    \centering
\begin{tabular}{|l|c|c|c|}
\toprule
\textbf{Criterion Removed} & \textbf{MRR} & \textbf{MAP} & \textbf{Recall} \\
\midrule
UNIQUE VALUES & 0.60 & 0.50 & 0.78 \\
INTERSECTION SIZE & 0.60 & 0.46 & 0.76 \\
JOIN SIZE & 0.59 & 0.49 & 0.77 \\
REVERSE JOIN SIZE & 0.59 & 0.49 & 0.80 \\
VALUE SEMANTICS & 0.59 & 0.49 & 0.78 \\
DISJOINT VALUE SEMANTICS & {0.65} & {0.56} & {0.80} \\
METADATA SEMANTICS & 0.54 & 0.46 & 0.78 \\
\midrule 
\textcolor{gray}{ALL PREFERENCES} & \textcolor{gray}{0.61} & \textcolor{gray}{0.49} & \textcolor{gray}{0.65} \\
\bottomrule
\end{tabular}
\end{table}

\begin{table}[]
    \centering
\caption{Performance comparison when the mentioned criterion is removed from the preferences in AutoJoin }
\label{table:autojoin_without_ranker}
   \begin{tabular}{|l|c|c|c|}
\toprule
\textbf{Preference Criterion Removed} & \textbf{MRR}  & \textbf{MAP}  & \textbf{Recall} \\
\midrule
UNIQUE VALUES &  0.31 & 0.31 & 1.0 \\
INTERSECTION SIZE & 0.30 & 0.30 & 1.0 \\
JOIN SIZE & 0.31 & 0.31 & 1.0 \\
REVERSE JOIN SIZE & 0.30 & 0.20 & 1.0 \\
VALUE SEMANTICS & 0.32 & 0.32 & 0.96 \\
DISJOINT VALUE SEMANTICS & 0.30 & 0.30 & 1.0  \\
METADATA SEMANTICS & 0.29 & 0.29 & 1.0 \\
\midrule
\textcolor{gray}{ALL PREFERENCES} & \textcolor{gray}{0.32} &  \textcolor{gray}{0.32} &  \textcolor{gray}{0.6}  \\
\bottomrule
\end{tabular}
\end{table}

Surprisingly, removing the disjoint semantics criterion leads to the an improvement in the search results for OpenData. We find that in OpenData lot of unrelated tables have columns with similar concepts. For example, disjoint semantics identifies that \texttt{County} column in~\cref{fig:topjoin_main}(b) and (d) are related. However, removing the disjoint value semantics drops the performance in AutoJoin dataset. This was also surprising as that dataset contains a lot of formatting variations, we expected the value semantics and the disjoint-value semantics to have same impact. \textbf{This shows the significance of disjoint value semantics as essential criteria for join-search specifically when the joins are fuzzy but no so much when the dataset contains a lot of related concepts.}

Another interesting observation in the AutoJoin ablation studies is that although the search performance when using only the value semantics is significant, removing only the value semantics does not hurt the performance. As the joinable columns rarely contain exact string matches, the two sets of disjoint value semantics (one that contains the values that are present in the query column but not in the candidate column and another that contains the values that are present in the candidate column and not in the query column) essentially represent the complete query and candidate columns. \textbf{Hence in case of low syntactic similarity, disjoint value semantic criterion acts similar to the value semantics criterion.}

Overall this study indicate that when the join task requires considerations of semantic relation between column values, the value semantics and disjoint value semantics are useful. When the dataset has additional information available for use, the metadata semantics is crucial. The unique values and size criteria play a more active role for relational databases, but less for data lakes. In fact, if one considers just the join size criterion for the GO Sales benchmark, that preference criterion by itself showed the same performance of WarpGate in our ablation studies (MRR: .68, MAP: .65, Recall: .99). 
This answers the research question (\ref{Q2:comparecriteria}).

\subsection{Embeddings}

One of the important considerations for the semantic joinability is the quality of embeddings. As discussed and emphasized in the DeepJoin and WarpGate papers~\cite{DeepJoin_Dong0NEO23,WarpGate_CongGFJD23}, embeddings from a finetuned model are more effective than the pretrained embeddings. 
However, it is not always feasible and ethical to use client data to finetune a model. So we include a study to identify which pre-trained embeddings are more useful. To understand the significance of the embedding quality, we present a comparison of three DeepJoin variants in this section. One, \textbf{DeepJoin-ST}, that uses column embeddings generated using  BERT-based sentence transformer. Second, the \textbf{DeepJoin-FT} that uses the FastText embeddings. Third, the \textbf{DeepJoin} that is the DeepJoin model finetuned on OpenData. We focus on DeepJoin as it seemed to perform better than the other embedding-based approach (WarpGate) for Fuzzy Join datasets.

\begin{figure}[!h]
  \centering
  \begin{subfigure}{0.7\columnwidth}
    \centering
    \inputtikz{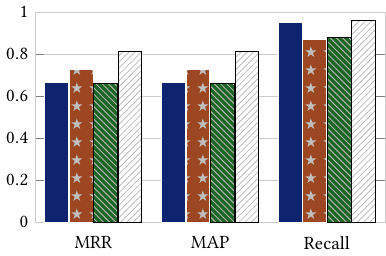}
    \vspace{-15pt}
    \caption{Valentine.}
    \label{fig:deepjoin_valentine}
  \end{subfigure}
  \begin{subfigure}{0.7\columnwidth}
    \centering
    \inputtikz{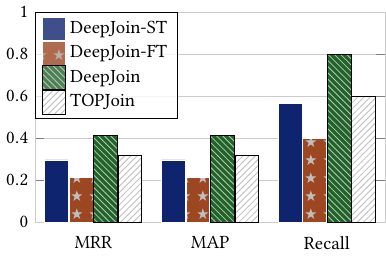}
    \vspace{-15pt}
    \caption{AutoJoin.}
    \label{fig:deepjoin_autojoin}
  \end{subfigure}
    \begin{subfigure}{0.7\columnwidth}
    \centering
    \inputtikz{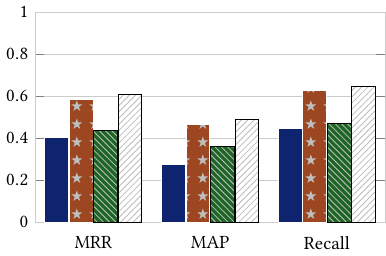}
    \vspace{-15pt}
    \caption{OpenData.}
    \label{fig:deepjoin_opendata}
  \end{subfigure}
    \begin{subfigure}{0.7\columnwidth}
    \centering
    \inputtikz{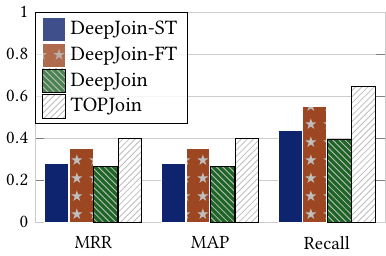}
    \vspace{-15pt}
    \caption{\nextia.}
    \label{fig:deepjoin_nextia}
  \end{subfigure}
  \caption{Comparison of different embeddings for columns. %
  DeepJoin-FT uses the FastText embedding, DeepJoin-ST uses a sentence transformer for column embeddings, DeepJoin is a finetuned model on OpenData.}
  \label{fig:deepjoin_embedding}
\end{figure}

While the FastText column embedding (in DeepJoin-FT) is the average of individual cell value embedding, the other two (DeepJoin-ST and DeepJoin) computes column embedding by considering the column text as a sentence. Given the shorter context window of the sentence transformer ($512$ tokens $\approx2500$ characters), the sentence embedding does not represent all the cell values. So we expect the sentence transformer embeddings to not perform as well as the FastText embedding. 
~\cref{fig:deepjoin_embedding} presents the results on 4 datasets; we also include~\ourmethod{} here for comparison. 
FastText embedding seems better than the Sentence Transformer for most of the cases; except the AutoJoin dataset (\cref{fig:deepjoin_autojoin}). 
Notably, the approach of generating embeddings for each cell value is impractical for large-scale data lakes, as it would be computationally expensive and inefficient, making it a non-viable solution for real-world product applications. Hence, we recommend using the sentence transformer approach.

\section{Summary and Guidelines}

Based on our comprehensive evaluation of join search methods and criteria across diverse benchmarks, we summarize the following practical guidelines for selecting appropriate techniques for different data contexts:

\begin{itemize}
    \item \textbf{Equi-joins} are unreliable when tables contain many integer keys with overlapping identifiers, but effective when semantically related columns exhibit high value overlap and unrelated ones do not.
    \item \textbf{Metadata} (e.g., column and table names) substantially improves join search performance when available.
    \item \textbf{Value semantics} is the most informative criterion for ranking join candidates.
    \item \textbf{Disjoint value semantics} aids fuzzy joins by filtering candidates with non-overlapping yet semantically related values, though its utility declines with highly interrelated concepts.
    \item \textbf{Uniqueness and join size} are valuable for relational databases but less so for heterogeneous data lakes.
    \item \textbf{Embedding models:} fine-tuning yields the best performance; otherwise, sentence-transformer embeddings outperform FastText.
    \item \textbf{Multi-criteria optimization} consistently outperforms any single criterion for context-aware joins.
\end{itemize}

\section{Conclusion}

We presented a comprehensive experimental analysis of joinable column discovery methods across diverse datasets and data characteristics. 
Our study examined how different criteria and approaches—ranging from syntactic and semantic joins to ensemble-based combinations—affect the effectiveness of context-aware join discovery. 

We curated seven benchmarks spanning enterprise databases, academic datasets, and open data lakes with manually annotated ground truth. 
Across one equi-join, two semantic-join, and one ensemble-based approach, we found that combining multiple joinability criteria yields more reliable performance across heterogeneous data environments.

Among the six key criteria we analyzed, metadata semantics, value semantics, and disjoint value semantics proved most impactful, underscoring the importance of combining structural and contextual cues. 
In the absence of fine-tuning data, sentence-transformer embeddings offer a strong alternative to fine-tuned models for semantic joins.

This study provides actionable insights into the strengths and limitations of existing methods and practical guidance for selecting effective techniques and criteria when building scalable, context-aware join discovery systems. 
To support further research and reproducibility, the code and experimental artifacts, along with the publicly shareable datasets, are available at \url{https://ibm.biz/context-aware-join}.

\section*{Artifacts}

All code, scripts, and experimental resources used in this paper are publicly available to support transparency and reproducibility.  
The full implementation of the evaluated methods, along with scripts to reproduce the experiments and figures, is available on GitHub at  
\url{https://github.com/IBM/ContextAwareJoin}.  
The repository also provides prepared datasets derived from prior public benchmarks, including the newly curated and manually annotated OpenData tables for evaluating context-aware join discovery, as well as pointers and citations to the original datasets, under  
\url{https://github.com/IBM/ContextAwareJoin/tree/main/datasets}.  
The manually annotated OpenData resources are additionally archived on Zenodo at  
\url{https://zenodo.org/records/15881731}.  
Due to confidentiality restrictions, the CIO database used in one of the benchmarks cannot be shared.  
All other datasets and instructions for reproducing the results are openly accessible, and the shared artifacts are released under open licenses for academic and research use.

\FloatBarrier

\balance

\bibliographystyle{IEEEtran}
\bibliography{biblio}

\end{document}